\documentclass[%
aip,
rsi,
amsmath,amssymb,
reprint,%
]{revtex4-1}

\usepackage{graphicx}
\usepackage{dcolumn}
\usepackage{bm}

\usepackage[utf8]{inputenc}
\usepackage[T1]{fontenc}
\usepackage{mathptmx}
\usepackage{etoolbox}

\makeatletter
\def\@email#1#2{%
\endgroup
\patchcmd{\titleblock@produce}
{\frontmatter@RRAPformat}
{\frontmatter@RRAPformat{\produce@RRAP{*#1\href{mailto:#2}{#2}}}\frontmatter@RRAPformat}
{}{}
}%
\makeatother
\begin{document}

\preprint{}

\title[Flat-top pulse NMR]{NMR measurements in dynamically controlled field pulse}
\author{Y. Ihara}
\email{yihara@phys.sci.hokudai.ac.jp}
\author{K. Hayashi}%
\affiliation{Department of Physics, Faculty of Science, Hokkaido University, Sapporo 060-0810, Japan.}
\author{T. Kanda}
\author{K. Matsui}
\author{K. Kindo}
\author{Y. Kohama}
\affiliation{Institute for Solid State Physics, The University of Tokyo, Kashiwa, Chiba 277-8581, Japan}%

\date{\today}

\begin{abstract}
We present the architecture of the versatile NMR spectrometer with software-defined radio (SDR) technology and its application to the dynamically controlled pulsed magnetic fields. 
The pulse-field technology is the only solution to access magnetic fields greater than 50 T, but the NMR experiment in the pulsed magnetic field was difficult because of the continuously changing field strength. 
The dynamically controlled field pulse allows us to perform NMR experiment in a quasi-steady field condition by creating a constant magnetic field for a short time around the peak of the field pulse. 
We confirmed the reproducibility of the field pulses using the NMR spectroscopy as a high precision magnetometer. 
With the highly reproducible field strength we succeeded in measuring the nuclear spin-lattice relaxation rate $1/T_1$, which had never been measured by the pulse-field NMR experiment without dynamic field control. 
We also implement the NMR spectrum measurement with both the frequency-sweep and field-sweep modes and discuss the appropriate choice of these modes depending on the magnetic properties of sample to be measured. 
This development, with further improvement at a long-duration field pulse, will innovate the microscopic measurement in extremely high magnetic fields. 
\end{abstract}

\maketitle

\section{\label{sec:intr}Introduction}

NMR spectroscopy is a powerful technique to explore the electronic state in materials from a microscopic viewpoint. 
By using the nuclear spins as the local magnetic probes, we can observe the microscopic magnetism of electrons around the nuclear sites through the hyperfine interactions between the nuclear and electronic spins. 
In spite of the microscopic nature of probing method the experimental setup around the sample is as simple as just mounting a sample in a radio-frequency (RF) coil. 
This simple geometry allows us to perform NMR experiments in extreme conditions such as low temperatures of a few mK \cite{yamashita-PRB102} and high pressures up to 9 GPa. \cite{kitagawa-JPSJ79} 
High magnetic field is another extreme condition of our interest. 
In high magnetic fields, increase in the nuclear magnetization improves the NMR signal intensity and 
increase in the electronic magnetization contributes to a large NMR frequency shift by creating larger internal fields at the nuclear site, which results in a better frequency resolution of the NMR spectrum. 
To take these advantages, NMR spectrometer for high field has been intensively developed and the magnetic fields available for high-resolution NMR spectroscopy is now stronger than 20 T.\cite{nagai-CRY41, hashi-JMR256}

\begin{figure}
\includegraphics[width=8.5cm]{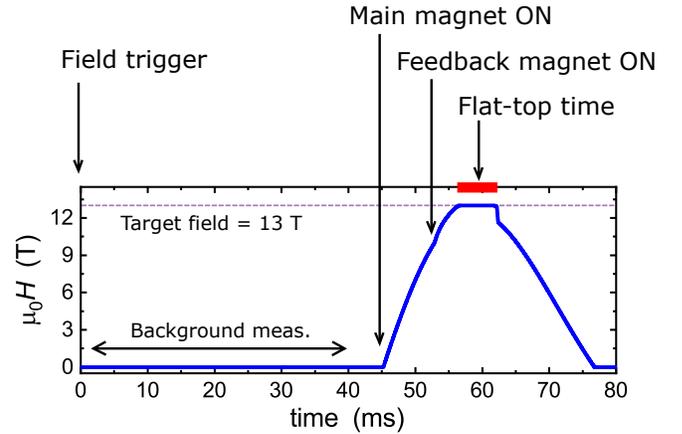}%
\caption{
Field profile of the dynamically controlled field pulse. 
NMR measurements are performed during the flat-top time, where field strength becomes constant by the PID control. 
To subtract the background voltage on the pickup coil precisely, we took background measurement time just before the onset of field pulse. 
}
\label{fig1} 
\end{figure}

NMR experiment in higher magnetic fields is crucial for the study of material science and fundamental solid state physics, 
because extremely high fields often bring the electronic spins into a nontrivial quantum state. 
The magnetic fields greater than 20 T are generated in a steady state by the hybrid magnet technology (up to 45.5 T) \cite{miller-IEEE13, pugnat-IEEE28, hahn-Nature570} or high-$T_c$ wire technology (up to 30.5 T) \cite{awaji-IEEE24,michael-IEEE29}. 
To access much higher magnetic fields, we use the pulse-field technology, with which fields of roughly 100 T are generated for the short duration of less than 1 second. \cite{jaime-PNAS109}
In spite of the enormous field strength, NMR measurement in pulsed magnetic fields has been a challenge because the resonant frequency changes with time following the continuously changing magnetic fields. 
Nevertheless several trials have been performed\cite{haase-SSNMR23,haase-AMR27,kozlov-SSNMR28,zheng-JPSJ78,meier-RSI83,stork-JMR234} and high quality results are available recently.\cite{orlova-PRL118, tokunaga-PRB99} 
So far, it was only possible to perform the field-sweep NMR spectrum measurement 
since the field pulses generated by the passive LCR circuits were not under control. 
To extract other physical quantities, such as nuclear spin-lattice relaxation rate $1/T_1$, from the pulse-field NMR experiment, the field pulse should be dynamically controlled (Fig.~\ref{fig1})\cite{kohama-RSI86}. 
Here we report the successful measurements of NMR spectrum and $1/T_1$ in the dynamically controlled field pulse (flat-top pulse) through the development of versatile NMR spectrometer. 
These apparatuses enable us to further improve the data quality and pioneer novel electronic states that appear in the extremely high magnetic fields.

\section{Measurement Setup}
\subsection{Dynamic field control}
In this study, we utilized for the first time the actively controlled flat-top pulse \cite{kohama-RSI86} to perform the pulse-field NMR experiment.  
The field profile and time schedules for a flat-top pulse is shown in Fig.~\ref{fig1}. 
To dynamically control the field strength during the field pulse, we insert a small feedback coil in the main magnet. 
The main magnet is driven by a capacitor bank (portable 2 kV, 15 mF at Hokkaido Univ. or built-in 10 kV 18 mF at ISSP) and the small feedback coil is driven by up to four 12 V batteries connected in series. 
The current on the feedback coil is controlled by the feedback voltage applied to the gate input of an insulated-gate bipolar transistor (IGBT) module. 
The magnetic fields at the sample space are measured by the induction voltage from the pickup coil wound at the end of the NMR probe. 
The pickup voltage is read at a sample rate of 1 MS/s by the analog to digital converter equipped with the multifunction reconfigurable I/O device USB-7856R (NI-National Instruments). 
Then, USB-7856R calculates the feedback voltage at the on-board field-programmable gate array (FPGA) device following the standard proportional-integral-derivative (PID) protocol and outputs voltage through the digital to analog converter. 
To irradiate RF pulses at an appropriate time during PID control (flat-top time) the NMR spectrometer should react to the field trigger at $t=0$. 
The general purpose in/out (GPIO) of USB-7856R receives an external trigger and generates trigger signal to the NMR spectrometer. 
By receiving the external trigger once at the USB-7856R, the time counter of USB-7856R can be precisely synchronized to the counter in the NMR spectrometer. 
Also GPIO can provide a field trigger signal to the capacitor bank together with the one for the NMR spectrometer if the capacitor bank needs an external trigger to generate a field pulse in the main magnet.

\begin{figure}
\includegraphics[width=8cm]{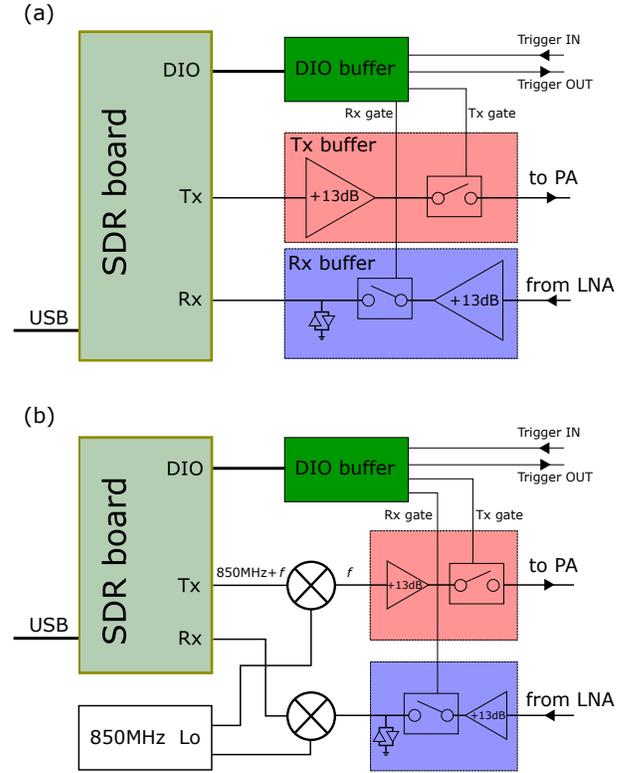}%
\caption{
Block diagrams of the SDR-based NMR spectrometers. 
The main SDR board constructs an RF network with Tx/Rx and generates/receives digital signals at DIO. 
The frequency band is (a) 100 MHz to 1 GHz and (b) 15 MHz to 250 MHz. 
For the high frequency measurement Tx output and Rx input are directly connected to the buffer boards. 
To decrease the measurement frequency lower than 100 MHz, frequency mixers are installed. 
With this option the Tx output is downconverted and Rx input is upconverted using the LO signal of 850 MHz. 
}
\label{fig2} 
\end{figure}

\subsection{Versatile NMR spectrometer with SDR technology}
We have developed an NMR spectrometer which can be flexibly optimized for the NMR measurement regardless of the type of electromagnets, i.e. steady or pulsed fields. 
Since the typical time window of the flat-top time is shorter than 100 ms, NMR measurements should be conducted at a precisely controlled timing and at a high repetition rate. 
We also need to implement a sophisticated measurement sequence such as rapid frequency skip at each scan. 
To accomplish this, we took advantage of the versatility of the software-defined radio (SDR) technology. 
The main SDR board is USRP-2901 (NI), which covers the frequency range from 70 MHz up to 6 GHz. 
The broad frequency band achieved by the SDR technology is difficult to cover with an ordinary analog-type heterodyne spectrometer. 
Our ultimate goal is to apply this method to the pulsed magnetic fields greater than 20 T, which is not easily accessible with the steady fields generated by superconducting magnets, 
and thus the frequency range required for the NMR experiment is normally higher than 100 MHz. 
Therefore, we use the direct RF input/output of the USRP-2901. 
We limit the lower frequency band to 100 MHz, as the linearity of the RF signal deteriorates at lower frequencies. 
To protect the SDR board from the over-voltage input, we inserted the transmission (Tx) and reflection (Rx) buffers, on which an RF switch, a fixed gain amplifier (ADL5536; Analog Devices), and 
back-to-back diodes for voltage clamping are mounted [Fig.~\ref{fig2}(a)]. 
At present, the specifications of amplifiers on the buffer boards limit the upper frequency band to approximately 1 GHz. 

In order to enable the measurements at frequencies lower than 100 MHz, we constructed a frequency conversion interface using a local oscillator signal (LO) of 850 MHz as shown in Fig~\ref{fig2}(b). 
For an analog system the quadrature detection is executed at a fixed frequency to maintain the orthogonality of inphase (I) and quadrature (Q) signals. 
In contrast to an analog system, our system uses a variable frequency for the quadrature detection, but a fixed one for LO. 
This is possible only with the broadband SDR technology, which reasonably maintains the I/Q orthogonality at any frequencies. 
With this low-frequency option, we can perform NMR measurement at $15 \sim 250$ MHz, which covers most NMR experiments in steady fields. 

The USRP-2901 is controlled through the USRP Hardware Drive (UHD) software (Ettus Research), which is the free and open-source software driver for USRP platform. 
The required Tx waveform is set in the personal computer (PC) and sent to the USRP-2901 through USB 3.0. 
We can easily modulate the baseband signal by programming the Tx waveform at PC, which enables the rapid frequency skip and irradiation of shaped pulse as we will discuss in \S IV-A. 
The timing of Tx generation is scheduled using the internal counter of URSP-2901. 
Since the field trigger coming from USB-7856R sets the counter to zero, timing for Tx generation is precisely synchronized to the pulse-field profile. 
The Rx signal is sampled at a rate of 10 MS/s with the vertical resolution of 12 bits. 
As the obtained data are continuously transferred to the PC, recording for a long time more than 1 second is possible. 
Synchronized with the Tx and Rx time, 
the digital output sends the gate signals to the power amplifier (PA) and the low-noise amplifier (LNA) to reduce the background noise and to protect small-power devices from a large signal, respectively. 
Here again, we placed a digital I/O (DIO) buffer to protect the SDR board and to provide sufficient drive current to the output signals.

\section{NMR measurement with flat-top pulse fields}

\begin{figure}
\includegraphics[width=7cm]{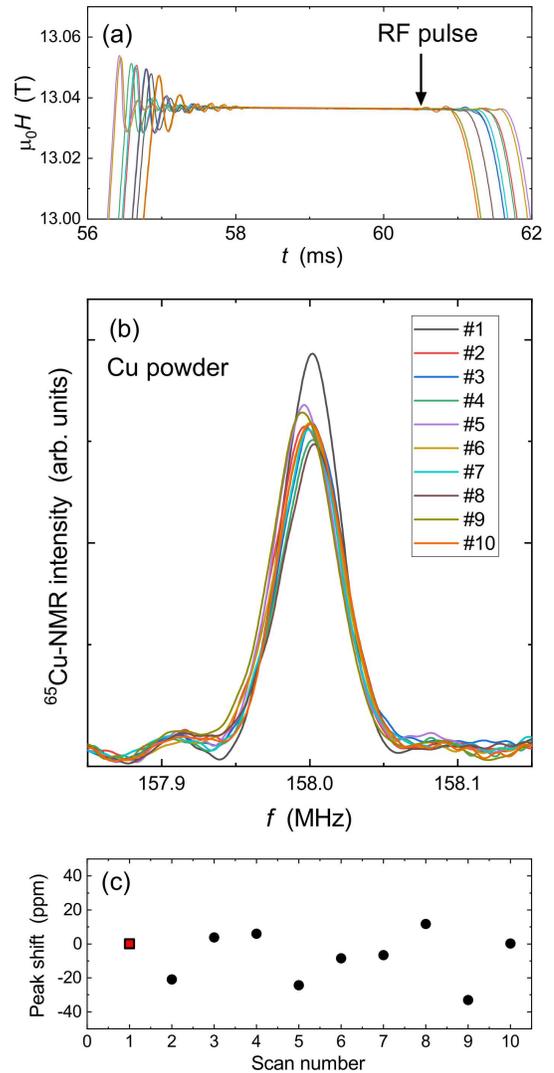}%
\caption{
Reproducibility test of 10 independent field pulses. 
(a) Field profiles of all field pulses around the flat-top time. 
The duration of flat-top time changes with the variation of peak fields generated by the main magnet. 
Even with this variation the magnetic fields are locked to the target value during the flat-top time. 
The downward arrow represents a time when the $^{65}$Cu-NMR measurement was performed. 
(b) $^{65}$Cu NMR spectra obtained at each field pulse and at a fixed carrier frequency. 
The magnetic fields are measured precisely from the peak positions of these FT spectra. 
(c) The peak position of $^{65}$Cu-NMR spectra for each field pulse. 
The distribution of magnetic fields with respect to the first field pulse (square) is less than 40 ppm. 
}
\label{fig3} 
\end{figure}

\subsection{Reproducibility of magnetic fields and NMR signals}

In a conventional field pulse generated by a passive LCR circuit, the magnetic field strength evolves continuously, thus a thermodynamic equilibrium state has never been achieved. 
This feature is not suitable for the measurement of thermodynamical properties, such as $1/T_1$, that requires to follow the relaxation process at certain condition. 
In contrast, with our flat-top pulse, magnetic field strength is dynamically controlled and tuned to a target field for the time duration of up to 15 \% of the total width of the field pulses. 
The NMR relaxation processes can be obtained within this flat-top time. 
Another important advantage is the high reproducibility of the field strength during the flat-top time. 
Without the dynamic control the maximum fields change with the magnet temperature even if the charge voltage is precisely controlled. 
To demonstrate the reproducibility of flat-top pulse, we generated 10 independent field pulses and their field profiles measured by the pickup coils are shown in Fig.~\ref{fig3}(a). 
To measure the absolute values of the external fields, we observed the $^{65}$Cu-NMR spectrum by using the apparatus shown in Fig.~\ref{fig2}(a). 
As the NMR frequency is proportional to the external fields, we can determine the magnetic field precisely from the peak frequency. 
The RF pulses for the $^{65}$Cu-NMR measurement are irradiated at $t=60.5$ ms, as pointed by a downward arrow in Fig.~\ref{fig3}(a). 
The carrier frequency is fixed to 157.95 MHz. 
The free induction decay (FID) signals after a single RF pulse were collected and their Fourier transform (FT) spectra are shown in Fig.~\ref{fig3}(b). 
The peak frequencies of each spectrum were determined by the Gaussian fitting and displayed in Fig.~\ref{fig3}(c), where the vertical axis is the deviation of the peak frequency from the value for the first pulse plotted in ppm scale. 
This result evidences that the field reproducibility is better than 40 ppm. 
We note here that the standard deviation of the integrated NMR intensity was calculated to be 6.7 \%, which is sufficiently small for a single scan measurement. 
This result is important to measure the relaxation time as we will discuss in the next section. 

\begin{figure}
\includegraphics[width=7cm]{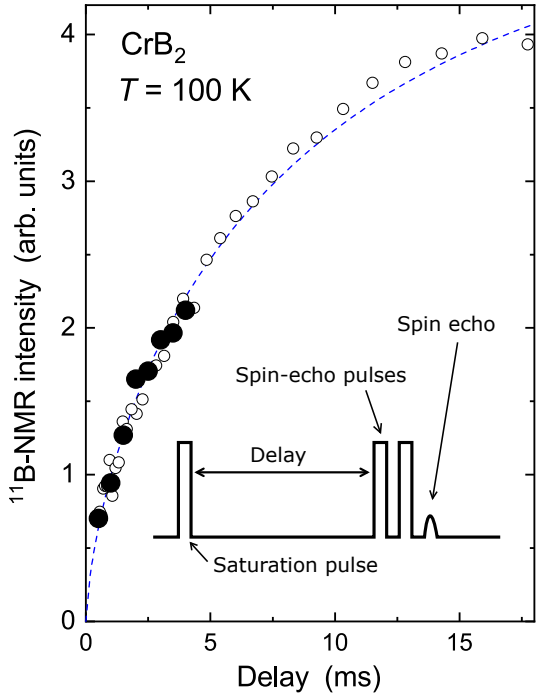}
\caption{
Relaxation profile of the nuclear magnetization measured for an antiferromagnet CrB$_{2}$ at $T=100$ K. 
The $^{11}$B-NMR intensity at each delay after the saturation pulse is recorded in pulsed fields (filled symbol) and steady fields (open symbols). 
Inset shows the RF pulse sequence for the relaxation rate measurement. 
The blue dashed line shows the result of least square fitting by the stretched exponential function. 
}
\label{fig4} 
\end{figure}

\subsection{Nuclear spin-lattice relaxation rate measurement}
To measure the relaxation profile of the nuclear magnetization after the saturation pulse, NMR signal intensity is recorded at each delay between the saturation and spin-echo pulses. 
The inset of Fig.~\ref{fig4} shows the RF pulse sequence for one scan. 
Since the spin-echo pulses disturb the free relaxation of the nuclear magnetization, only one point in the relaxation profile can be measured by a single scan. 
For the pulse-field NMR measurement, we need to measure the NMR intensity for each delay at independent field pulses. 
Therefore, the reproducibilities of magnetic field and NMR intensity are crucially important. 
As we confirmed the reproducibility of NMR signal intensity for independent field pulses, we can now measure the relaxation profile by repeating the NMR measurements with different delays. 

To demonstrate the $1/T_1$ measurement, we measured the relaxation profile of a typical antiferromagnet CrB$_{2}$ at 100 K. 
This compound shows an antiferromagnetic phase transition at $T_{\rm N}= 88$ K, \cite{barnes-PLA29, castaing-SSC7} and thus $T_1$ is reasonably short near the transition temperature. \cite{kitaoka-JPSJ49}
We measured the $^{11}$B-NMR signal at a target field of 13.0 T, which corresponds to the NMR frequency of approximately 178 MHz. 
In Fig.~\ref{fig4}, the results collected in the pulsed fields are shown by the filled symbols. 
We repeated the NMR scans twice for each delay to improve the data accuracy. 
The longest delay was 4.0 ms, which is limited by the width of the field pulse for the present experiment. 
As a reference, we plotted the relaxation profile measured in a steady field of 13.0 T by the empty symbols. 
We confirm that the results of pulse-field NMR measurement perfectly follow those in the steady fields up to the maximal delay possible for the present flat-top time. 

The relaxation profile $M(t)$ is fitted with a stretched exponential function 
\begin{equation}
M(t) =M_0\left[1-A\exp \left( -\left(\frac{t}{T_1}\right)^{\beta}\right) \right]. 
\end{equation}
Here, $M_0$ and $A$ are the nuclear magnetization in the thermal equilibrium and the saturation coefficient, respectively. 
We introduce a stretched exponent $\beta$ to better fit the experimental results. 
From the fit we obtained $T_1 =8.9 $ ms and $\beta = 0.7$ and the resulting relaxation curve is plotted by the blue dashed line. 
Although this experiment demonstrates the validity of $1/T_1$ measurement in the pulsed magnetic fields and opens a possibility to measure $1/T_1$ at extremely high fields, 
the maximum delay is still too short to fit the overall relaxation profile. 
As a next step, we should perform the NMR measurement in flat-top pulse with a long-duration pulse magnet that has a pulse width exceeding 1 second. \cite{herlach-RPR62, matsui-RSI92}
We note that the ability to measure $1/T_{1}$ is easily expanded to the nuclear spin-spin relaxation rate $1/T_2$ measurement, which requires much less duration time of typically a few milliseconds.

\section{NMR spectrum measurement for broad spectrum}

\begin{figure}
\includegraphics[width=8cm]{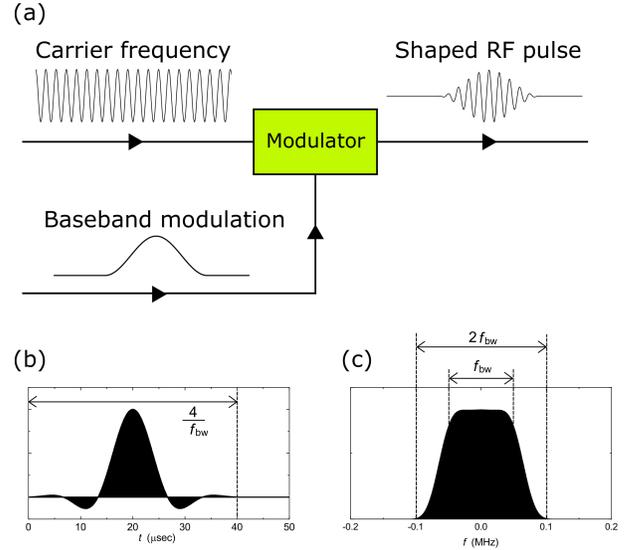}
\caption{
(a) Schematic diagram for the baseband modulation. 
The continuous carrier frequency is mixed with the baseband modulation waveform to construct a shaped pulse. 
The baseband waveform of the ham-flat RF pulse (b) and its FT power spectrum (c). 
The bandwidth $f_{\rm bw}$ is set to 100 kHz. 
}
\label{fig5} 
\end{figure}

When the NMR spectrum is narrower than the frequency window for a single RF pulse, which is typically a few hundreds of kHz, entire NMR spectrum can be measured at a fixed field and frequency as in the case of $^{65}$Cu-NMR spectra in Fig.~\ref{fig3}(b). 
For broader NMR spectra, however, NMR signal intensity should be recorded during either frequency sweep at a fixed field or field sweep at a fixed frequency. 
In a steady field, we choose one of these measurement modes depending on the overall spectrum width and physical properties at the measurement fields. 
In the previous pulse-field NMR studies, only the field-sweep experiment was performed. \cite{zheng-JPSJ78, stork-JMR234, orlova-PRL118, tokunaga-PRB99}
Here we adopt both modes using the dynamically controlled field pulse. 

\begin{figure}
\includegraphics[width=8cm]{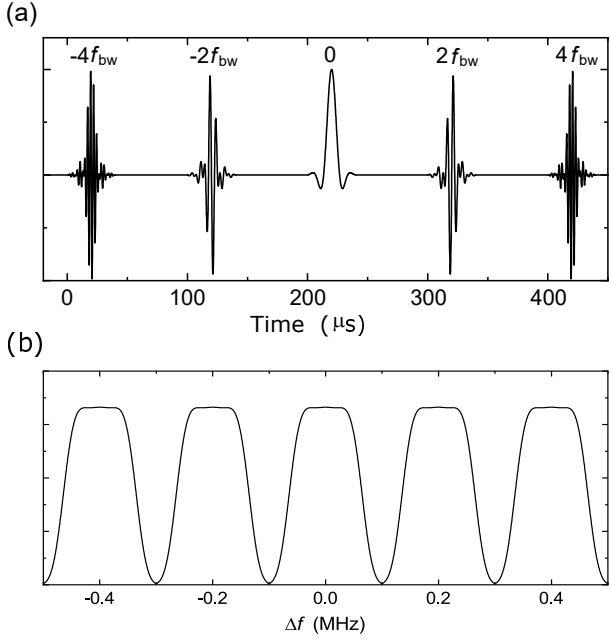}
\caption{
(a) The RF pulse sequence for the frequency-sweep mode during the flat-top time period and FT power spectra of each RF pulse. 
Wide range frequency sweep is realized within a short time, while the overlap of Tx power spectrum is minimized to avoid the saturation of nuclear magnetization. 
Here single RF pulse for each frequency is displayed for simplicity. 
In reality spin-echo pulses sequence, such as $\pi/2-\tau-\pi$ sequence, is generated at each Tx frequency.
}
\label{fig6} 
\end{figure}

\subsection{Frequency-sweep mode with flat-top pulse}
We perform the frequency-sweep experiment during the flat-top time period in almost the same way as in the steady field. 
The only and the most different point is that we need to sweep the measurement frequency in a short time, namely less than 1 ms. 
If we irradiate several RF pulses at the same frequency in one field pulse, the signal intensity would gradually diminish because of the saturation of the nuclear magnetization. 
To avoid this, we should irradiate only one RF pulse sequence for each frequency by shifting the Tx frequency rapidly during the flat-top time. 
Although high speed frequency skip is rather difficult to execute with the ordinary analog heterodyne structure, the versatility of SDR board enables rapid and precise frequency shifting by its digital baseband modulation feature, 
with which the modulation of carrier frequency by the baseband signal [Fig.~\ref{fig5}(a)] is performed at a digital signal processing circuitry. 
Then, the Tx frequency shift of $\Delta f$ is achieved by multiplying a phase factor of $\phi(t)=\exp (2 \pi i \Delta f t)$ to the carrier frequency. 

Moreover, as rectangular RF pulses will irradiate the RF power in a broad frequency range, which is characterized by the sinc function with several side lobes, we employed a shaped RF pulse to further avoid the irradiation to the unwanted frequency. 
The baseband-modulation waveform of the shaped RF pulse is a convolution of the sinc function and the hamming window. (ham-flat)
\begin{equation}
w(t)= \left[ 0.54-0.46\cos \left( \frac{\pi f_{\rm bw}}{2}t\right) \right] \frac{ \sin \left(1.5 \pi \left( f_{\rm bw}t-2\right) \right)}{1.5 \pi \left( f_{\rm bw}t-2\right)}.
\end{equation}
Here, $f_{\rm bw}$ is a Tx frequency bandwidth. 
The waveform of the ham-flat pulse and its FT power spectrum are displayed in Figs.~\ref{fig5}(b) and (c). 
The FT spectrum shows that the RF power within the frequency window of $f_{\rm bw}$ is almost constant, which is characterized by the RF power at $\pm 0.5f_{\rm bw}$ being $-0.7$ dB. 
The RF power decays rapidly at lower or higher frequencies and becomes smaller than $-22$ dB at $\pm f_{\rm bw}$. 
By irradiating these shaped RF pulses at 2$f_{\rm bw}$ step as shown in Fig.~\ref{fig6}(a), only one RF pulse sequence is irradiated for each frequency without any overlap. 
The vacancies between two frequencies are filled by repeating the field-pulse generation with the initial frequency shifted by $f_{\rm bw}$. 
The flat-top pulse field is crucially important for the frequency-sweep mode firstly because the field should be fixed during the frequency sweep and also 
because the field strength has to be reproducible to fill the frequency vacancies. 
By repeating field-pulse generation, we can improve the signal to noise ratio (SNR). 

This frequency-sweep mode with rapid frequency skip can be used in the steady-field experiment to accelerate the repetition rate. 
When we use a single Tx frequency with a conventional NMR spectrometer, 
the repetition rate of the Tx irradiation is limited by the time scale of material-specific $1/T_1$, which is much longer than millisecond order. 
By using multiple Tx frequencies, for example five frequencies as shown in Fig.~\ref{fig6}(b), we can perform the NMR experiment at five different frequencies in parallel 
because these Tx frequencies do not interfere with each other. 
As a result, the repetition rate of Tx irradiation will become five times faster. 

\begin{figure}
\includegraphics[width=8cm]{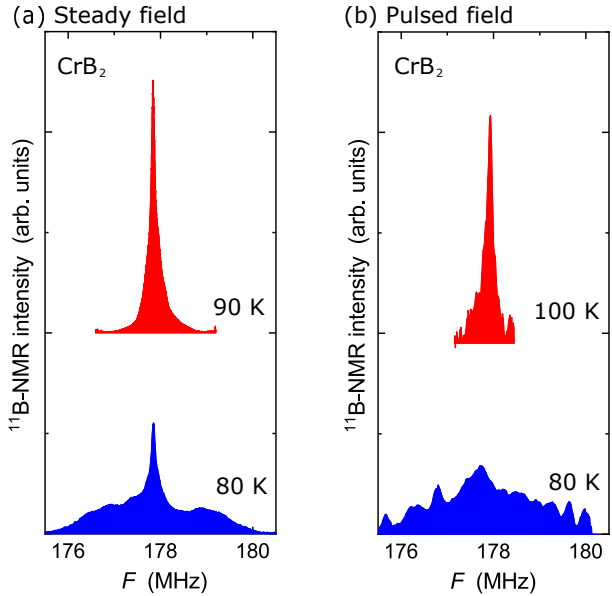}
\caption{
The frequency-sweep NMR spectra for CrB$_{2}$ above and below $T_N$. 
The NMR spectra were obtained in steady field (a) and pulsed field (b). 
The external magnetic field is 13.0 T for both cases. 
Sharp peak in the paramagnetic state and broadening in the ordered state were consistently observed in the pulsed field. 
}
\label{fig7} 
\end{figure}

As an example of the frequency-sweep spectrum measurement we show the $^{11}$B-NMR spectra for CrB$_{2}$ obtained in steady and pulsed fields in Figs.~\ref{fig7}(a) and (b). 
The target magnetic field is 13.0 T. 
A spectrum broadening was observed below $T_N= 88$ K because of the appearance of internal fields generated by the ordered moments. 
The sharp spectrum above $T_N$ and broadened spectrum below $T_N$ were both consistently observed in the pulse-field NMR measurement. 
We generated 40 field pulses with 3 RF pulses at each flat-top time to obtain the broadened NMR spectrum at 80 K. 
We repeated a lot of field-pulse generation to sweep over a broad frequency range and to increase the SNR as the signal intensity is significantly reduced below $T_N$. 
Nevertheless, the SNR of the NMR spectra obtained in the pulsed field cannot be compared to those in steady fields especially for the broad spectrum. 
Since the present frequency window covered by one field pulse is limited by approximately 1 MHz by the quality factor of the RF tank circuit, 
we need to repeat the field-pulse generation for many times to observe the full spectral shape. 
In the case of broad NMR spectrum, a field-sweep mode with long-duration pulse fields is more appropriate as we explain in the next section.

\begin{figure}
\includegraphics[width=8cm]{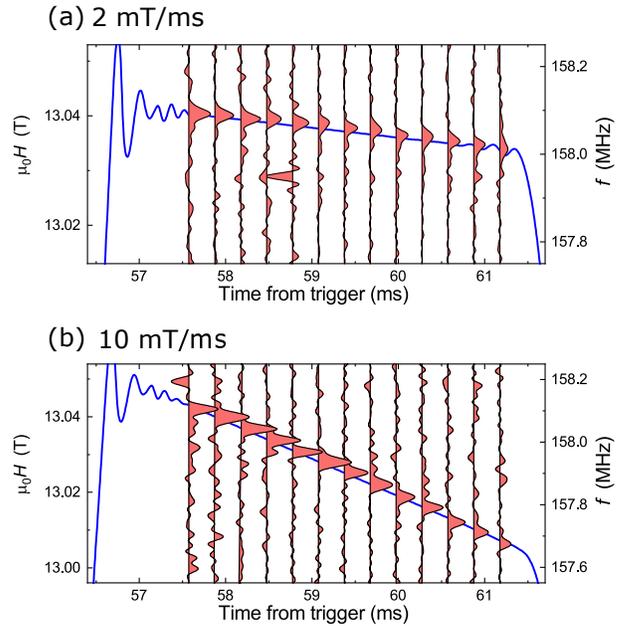}
\caption{
The field profile of the slope-top pulses (blue line). 
The constant sweep rate is (a) 2 mT/ms and (b) 10 mT/ms. 
Red peaks are the FT NMR spectra of $^{65}$Cu FID signals measured at corresponding time. 
Since the NMR measurement is one of the most precise magnetometer constant and arbitrary rate field-sweep ability is clearly demonstrated. 
}
\label{fig8} 
\end{figure}

\subsection{Field-sweep mode with slope-top pulse}
As an alternative measurement mode for broad NMR spectra, we sweep the magnetic fields during the irradiation of RF pulses at a fixed frequency. 
This field-sweep mode is frequently used for very broad spectrum as the sweeping range is not limited by the mechanical parameters of the RF tank circuit. 
The field-sweep experiment has been already realized with the field pulse without the PID feedback control. \cite{zheng-JPSJ78, stork-JMR234, orlova-PRL118, tokunaga-PRB99}
However, the relative NMR intensity is modified by the continuously changing sweep rate for the half-sinusoidal field-pulse profile. 
Here, we sweep the magnetic fields by changing the target field at a constant rate during the PID control (slope-top pulse). 
Fig.~\ref{fig8} shows the resulting magnetic field profiles at two sweep rates, 2 mT/ms and 10 mT/ms. 
The field profiles shown by the blue solid lines were measured by the pickup coil. 
To confirm the field strength at each moment, we performed the $^{65}$Cu-NMR measurement at every 0.3 ms starting from 57.5 ms. 
The peak positions of the $^{65}$Cu-NMR spectra follow the blue lines for both sweep rates, evidencing that the field strength is nicely controlled to decrease at a constant rate. 
We used a small RF power to avoid saturation of nuclear magnetization even at a very fast repetition rate. 
With this method suppression of NMR intensity was observed only for the last few NMR spectra around 61 ms, although $T_1$ of $^{65}$Cu at 77 K is longer than 10 ms. 
Since the external fields change at a constant rate, NMR spectrum intensity is correctly measured by irradiating the RF pulse at a fixed repetition rate. 

The main magnet we used for this experiment generates a field pulse with the total width of $30$ ms. 
With this magnet a magnetic field of 13.0 T is PID controlled for 3.5 ms (57.5 ms to 61 ms). 
Therefore, when we use the sweep rate of 10 mT/ms, 
we can obtain the NMR spectrum in the field range of 35 mT by one field-pulse generation, which corresponds to 400 kHz for $^{65}$Cu nuclear spins and is narrower than the frequency window of the frequency-sweep mode. 
To perform the field-sweep experiment over a broader field range, we connect the data obtained in the other slope-top field pulse starting from the last field of the previous field pulse. 
However, as the optimization of the field-generation parameters is imposed each time by shifting the field window, present field-sweep range is not sufficient for the measurement of broad NMR spectra. 
The use of long-duration field pulse with the pulse width longer than 1 s permits to generate the slop-top field with the duration of $\sim 100$ ms. 
This will allow us to sweep magnetic fields for 1 T by a single field-pulse generation.

\subsection{Consideration for the choice of modes}

As a result of the development of the flat-top and slope-top pulses we can choose the measurement modes of NMR spectra depending on the target materials. 
Here we discuss the advantages and disadvantages for these modes which should be considered to choose the best measurement mode. 
When the overall spectral width is broader than a few MHz, the field-sweep mode with the long-duration field pulse is the first choice. 
A disadvantage of this setup is a long cooling time of the long-duration pulse magnet, which typically takes a few hours. 
In this respect, the frequency-sweep mode in a smaller pulse magnet with the pulse width of approximately 30 ms is a better choice when the NMR spectral width is narrower than 2 MHz. 
Within the spectral width, we can cover the full frequency range by few field pulses and increase the SNR by quickly repeating the field-pulse generation. 
The frequency-sweep mode with long-duration pulse field is not recommended as the frequency range measurable during one field pulse is limited by the quality factor of RF tank circuit 
and repeating the same frequency during one field pulse will reduce the NMR intensity by the saturation of nuclear magnetization. 
For the frequency sweep of 1 MHz, 10 RF pulses sufficiently cover the full frequency range, which can be generated within 3 ms by a repetition time of 0.3 ms. 
The long-duration pulse field should be used for the sample whose nuclear magnetization cannot be polarized during the field-pulse duration of 30 ms. 
In this case frequency-sweep measurement should be performed at the very end of the flat-top time to give ample time for the polarization of nuclear spins. 
Another case to select frequency-sweep mode is to measure the sample that shows a field-induced phase transition. 
Field sweep across the critical magnetic field results in the drastic change in the spectral shape at the middle of the NMR spectrum, 
and thus the entire spectral shape cannot be measured. 
Frequency-sweep mode at slightly above and below the critical magnetic field will clearly reveal the microscopic magnetism in the field-induced electronic state.

\section{Summary}
We demonstrated various operating modes of our SDR-based NMR spectrometer using the flat-top field pulse. 
We first confirmed the reproducibility of field strength for independent field pulses, which is crucial to improve the SNR, to measure relaxation profile of nuclear magnetization, and to measure frequency-sweep NMR spectrum. 
The $1/T_1$ and frequency-sweep NMR spectrum measurements, which were difficult with a field pulse without the dynamic control, were successfully performed in a metallic antiferromagnet CrB$_{2}$. 
We also developed the slope-top field pulse to enable the NMR spectrum measurement with field-sweep mode. 
These results open a possibility to measure microscopic magnetism in extremely high magnetic fields. 
However, since the present study was performed in the field pulse with approximately 30 ms time duration, flat-top time was not sufficiently long for some materials. 
To further expand the applicability of NMR experiment in high fields, these NMR technology should be transferred and adapted to the pulse magnet with longer pulse duration. 

\begin{acknowledgements}
One of the authors (K.M.) is a research fellow of the Japan Society for the Promotion of Science (JSPS). 
This work was partially supported by the JSPS Grant-in Aid for Scientific Research (Grant nos. 18H01163, 19H01832, and 20K20892), the Futaba Foundation, and ISSP Institutional Collaborative Research Program. 
\end{acknowledgements}

\section*{Data availability}
The data that support the findings of this study are available from the corresponding author upon reasonable request.

\end{document}